\documentclass[%
 aip,
 amsmath,amssymb,
 groupaddress,
 reprint,%
]{revtex4-1}

\usepackage{graphicx,siunitx}
\usepackage{dcolumn}
\usepackage{bm}
\usepackage[version=4]{mhchem}
\usepackage[mathscr]{euscript}
\usepackage[utf8]{inputenc}
\usepackage[T1]{fontenc}
\usepackage{mathptmx}
\usepackage{xcolor}

\newcommand{\half}{{\textstyle\frac{1}{2}}}

\newcommand{\kB}{k_{\mathrm{B}}}
\newcommand{\kT}{\kB T}
\newcommand{\rg}{r_{\mathrm{g}}}
\newcommand{\rd}{r_{\mathrm{d}}}

\newcommand{\df}{d_{\mathrm{f}}}
\newcommand{\dr}{{\mathrm d}r}

\newcommand{\RL}{R_{\mathrm L}}
\newcommand{\RS}{R_{\mathrm S}}

\newcommand{\pH}{p\ce{H}}
\newcommand{\mM}{\mathrm{mM}}

\newcommand{\rhod}{\rho_{\mathrm{d}}}
\newcommand{\rhop}{\rho_{\mathrm{p}}}
\newcommand{\phid}{\phi_{\mathrm{d}}}
\newcommand{\phip}{\phi_{\mathrm{p}}}

\newcommand{\etac}{\eta_{\mathrm{c}}}
\newcommand{\phic}{\phi_{\mathrm{c}}}
\newcommand{\betac}{\beta_{\mathrm{c}}}
\newcommand{\gc}{g_c} 
\newcommand{\etamax}{\eta_{\mathrm{max}}}

\newcommand{\Vfree}{V_{\text{free}}}
\newcommand{\myav}[1]{\langle{#1}\rangle}

\newcommand{\myvec}[1]{{\mathbf{#1}}}
\newcommand{\Ivec}{\myvec{I}}
\newcommand{\fvec}{\myvec{f}}
\newcommand{\rvec}{\myvec{r}}
\newcommand{\vvec}{\myvec{v}}
\newcommand{\rhatvec}{\hat{\rvec}}

\newcommand{\Eqref}[1]{Eq.\ \eqref{#1}}
\newcommand{\Eqsref}[1]{Eqs.\ \eqref{#1}}
\newcommand{\Figref}[1]{Fig.\ \ref{#1}}

\newcommand{\Refcite}[1]{Ref.\ \onlinecite{#1}}

\newcommand{\latin}[1]{{\itshape #1}}
\newcommand{\cf}{\latin{cf.\@}}

\newcommand{\ie}{\latin{i.e.\@}}
\newcommand{\etal}{\latin{et al.\@}}
\newcommand{\per}{\latin{per}}
\newcommand{\viz}{\latin{viz.}}

\newcommand{\german}[1]{{\itshape #1}}
\newcommand{\ansatz}{\german{ansatz}}

\begin{document}

\preprint{AIP/123-QED}

\title[2D depletion]{Diffusion, phase behavior and gelation in a two-dimensional layer of colloids in osmotic equilibrium with a polymer reservoir}

\author{Sam E. Griffiths}
\affiliation{School of Physics and Astronomy, The University of Edinburgh, Edinburgh EH9 3FD, United Kingdom
}%

\author{Nick Koumakis}%
\affiliation{School of Physics and Astronomy, The University of Edinburgh, Edinburgh EH9 3FD, United Kingdom
}%

\author{Aidan T. Brown}%
\affiliation{School of Physics and Astronomy, The University of Edinburgh, Edinburgh EH9 3FD, United Kingdom
}%

\author{Teun Vissers}
\affiliation{School of Physics and Astronomy, The University of Edinburgh, Edinburgh EH9 3FD, United Kingdom
}%

\author{Patrick B. Warren}
\affiliation{Hartree Centre, Science and Technology Facilities Council (STFC), Sci-Tech Daresbury, Warrington, WA4 4AD, United Kingdom}

\author{Wilson C. K. Poon}%
\affiliation{School of Physics and Astronomy, The University of Edinburgh, Edinburgh EH9 3FD, United Kingdom
}%

\date{\today}

\begin{abstract}
The addition of enough non-adsorbing polymer to an otherwise stable colloidal suspension gives rise to a variety of phase behavior and kinetic arrest due to the depletion attraction induced between the colloids by the polymers. We report a study of these phenomena in a two-dimensional layer of colloids. The three-dimensional phenomenology of crystal-fluid coexistence is reproduced, but gelation takes a novel form, in which the strands in the gel structure are locally crystalline. We compare our findings with a previous simulation and theory, and find substantial agreement. 
\end{abstract}

\maketitle

\section{\label{sec:intro}INTRODUCTION}
Adding non-adsorbing polymers to a suspension of hard-sphere colloids (radius $R$, volume fraction $\phic$) induces a depletion attraction between the particles.\cite{AO, Vrij1976, Lekker2011} Exclusion of polymers from the space between two nearby particles leaves an unbalanced osmotic pressure pushing them together. The depth and range of the depletion attraction $U_{\rm dep}(r)$ between two particles with centre-to-centre distance of $r$ is proportional to the polymer activity, $a_p$, and the polymer's radius of gyration, $\rg$, respectively. When the size ratio $\xi = \rg/R \lesssim 0.30 \pm 0.05$,\cite{cptheory, cpphaseexpt} the equilibrium $(\phic, a_p)$ phase diagram displays an expanded region of fluid-crystal (F-X) coexistence at $a_p \gtrsim O(10^{-1})$,\cite{PoonReview} which occurs for $0.494 < \phic < 0.545$ at $a_p = 0$. 

`Buried' within the equilibrium F-X coexistence region is a metastable gas-liquid (G-L) coexistence binodal that terminates at a critical point. A homogeneous sample inside this binodal should in principal first phase separate into metastable coexisting G-L phase, before further separating into equilibrium F-X coexistence.\cite{Evans1997} This scenario is, however, seldom observed, because kinetic arrest intervenes. 

For $0.1 \lesssim \phic \lesssim 0.3$, samples inside the metastable G-L coexistence region will phase separate by spinodal decomposition into a bicontinuous structure. With time, this structure coarsens in length scale, and the concentration difference between the two phases increases, until the concentration of the liquid phase crosses the `attractive glass transition'. line\cite{Pham2002} The texture arrests and the system becomes a gel.\cite{Zaccarelli2007}$^,$\footnote{Note that if the colloids are too polydisperse to crystallise, this scenario still holds, except that, now, G-L coexistence is no longer metastable, but is the sole equilibrium thermodynamic phase transition in the system.} 

Such `depletion gels' have been intensively studied for some time, but mysteries remain, perhaps especially how they age with time. In some cases, a depletion gel can undergo sudden gravitational collapse after an apparently quiescent period in which little seems to happen macroscopically.\cite{Piri1995, Harich2016, Graaf2019, Ramos2020} This and other aging phenomena are expected on thermodynamic grounds. A depletion gel is metastable. There is therefore a driving force for evolution towards the lowest free energy state, which is F-X coexistence.


What we have summarised so far pertains to bulk colloid-polymer mixtures. At first sight, there is little incentive to study  two-dimensional (2D) systems: presumably, any difference to bulk behavior would be merely quantitative. However, this intuition turns out to be incorrect.

\begin{figure}[t]
\centering
  \includegraphics[width=0.48\textwidth]{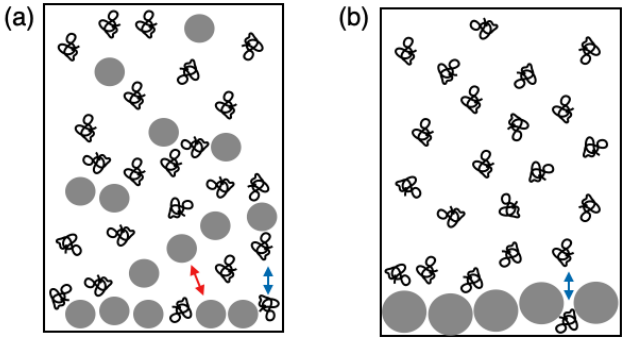}
  \caption{Different ensembles for 2D colloid-depletant mixtures, with the depletant represented as polymer coils, but they can also be smaller colloids or micelles. (a) Colloids with $z_0 \gg R$. Grand-canonical for both colloids and depletants in the wall layer (bottom),\cite{Dinsmore1997} with exchanges of both species with the bulk (blue and red double arrows). (b) Colloids with $z_0 \lesssim R$. Grand-canonical for the depletants in the surface layer, which exchange with the bulk (blue double arrow); but canonical for the colloids, which are all sedimented, with no exchange with the bulk.\cite{Hobbie1998, Dinsmore2012}}
  \label{fig:ensemble}
\end{figure}

An early 2D study extended the `primitive' theory for bulk phase behavior\cite{cptheory} to calculate the phase diagram of a bulk colloid-polymer mixture in the presence of a hard wall.\cite{Poon1994} The theory predicts that depletion-induced wall adsorption induces wall freezing (= crystallization) at depletant concentrations below the bulk F-X coexistence boundary. An interesting subtlety, \Figref{fig:ensemble}(a), is that both colloids and depletants at the surface are in osmotic equilibrium with the bulk, with which they can exchange both species. Thus, unlike in the bulk, which is a canonical ensemble where F-X coexistence is possible, surface crystallization occurs in a grand-canonical ensemble, so that the crystal fraction jumps from 0 to 1 at the critical bulk polymer concentration (at fixed colloid volume fraction). Experiments using large ($\RL = \SI{0.23}{\micro\meter}$) and small ($\RS = \SI{0.035}{\micro\meter}$) charge-stabilised polystyrene spheres (screening length $\approx \SI{5}{\nano\meter}$) in which the small spheres act as depletants~\cite{Dinsmore1997} confirmed this feature, and found reasonable agreement with theory for the wall crystallization boundary. 

The sedimentation height $z_0$ of the large spheres in these experiments, defined such that the number of particles of radius $R$ in a dilute suspension decreases with height according to $n(z) = n(0)e^{-(z-R)/z_0}$, is $z_0 = \SI{160}{\micro\meter} \gg \RL$, as assumed in the theory.\cite{Poon1994} Two other experiments used much lower  $z_0/R$. Savage \etal\ \cite{Dinsmore2012} used $R = \SI{0.7}{\micro\meter}$ polystyrene spheres ($z_0 \approx 8\,R$) with non-ionic surfactant micelles as depletants.  Hobbie\cite{Hobbie1998} studied binary polystyrene colloids with $\RL = \SI{1.45}{\micro\meter}$ ($z_0 \approx 0.4\,\RL$) and $\RS = \SI{0.107}{\micro\meter}$. In each case, gravity and the particle-wall depletion attraction  sufficed to confine all of the large particles to an effectively 2D layer (at low enough bulk concentration). 

Now, the surface layer is a canonical ensemble of a fixed number of large particles, but a grand-canonical ensemble of the smaller depletants, which freely exchanges with the bulk, \Figref{fig:ensemble}(b). In this `semi-grand canonical' scenario,\cite{Lekker1990} wall crystallization takes the form of F-X coexistence, with an increasing crystal fraction as the bulk concentration of depletants (which controls the surface depletant chemical potential) increases. This is indeed seen in experiments,\cite{Hobbie1998, Dinsmore2012} where crystal nucleation is two-stepped, proceeding via an intermediate gas-liquid phase separation, as predicted by theory.\cite{Evans1997}. Such kinetics is seldom seen in bulk, where the metastable gas-liquid critical point nevertheless enhances crystal nucleation.\cite{Wolde1997}  

Cerd\`a \etal\ have simulated a system mimicking a 2D colloid-polymer mixture: nearly-hard disks interacting via an Asakura-Oosawa (AO) potential, which is widely used to model polymer-induced depletion.\cite{Lekker2011} They worked at a size ratio $\xi = 0.1$, and probed the behavior at a surface colloid area fraction $\eta_{\rm c} = 0.157$. Cerd\`a \etal\ found F-X coexistence when the contact value of the AO potential, $U_0 \geq 3.130\,\kT$. At the even higher $U_0 = 7\,\kT$, they observed ramified clusters with fractal dimension $\df \approx 1.4$, the exponent for diffusion-limited cluster aggregation (DLCA). However, the local structure of the cluster strands is, unlike DLCA, clearly crystalline. This finding has yet to be confirmed by experiments. 

Here, we study a 2D colloid-polymer mixture in which we quantify how two-dimensionality by comparing diffusivities next to a wall against Fax\'en's prediction.\cite{Faxen1921, HB73, Chio2020} The equilibrium phase diagram is obtained and compared with theory.\cite{Poon1994} We confirm the locally-crystalline nature of the ramified clusters at the highest attractions.\cite{Cerda2004} Quantitatively, the time dependence of the average cluster size $\bar n \sim t^z$, with $0.5 \lesssim z \lesssim 0.6$. We discuss the possible origin of this exponent in the screened near-wall hydrodynamics of clusters, and propose why the local structure of our ramified clusters are crystalline, in striking contrast to the case of 3D depletion gels.\cite{Griffiths2017}

\section{MATERIALS AND METHODS}
We used \SI{1.5}{\micro\meter} diameter ($2R$) silica spheres (Bangs Laboratories, density $\rho \approx \SI{2}{\gram\per\cubic\centi\meter}$). The work we report is part of a larger program\cite{GriffithsPhD} studying the effect of self-propelled particles on colloidal gels, where the self-propelled particles are motile {\it Escherichia coli} bacteria,\cite{Linek2016} for which we need to know the behavior of the cell-free system. We therefore dispersed our colloids in a phosphate motility buffer (PMB) commonly used to study motile {\it E. coli} ($6.2\,\mM$ \ce{K2HPO4}, $3.8\,\mM$ \ce{KH2PO4} and $0.1\,\mM$ EDTA at $\pH\approx 7.5$). We removed the \ce{NaCl} usually included in standard PMB to limit the ionic strength to $I = 22.4\,\mM$ (screening length \SI{2}{\nano\meter}). Under these conditions, our colloids do not visibly aggregate, so that any residual interparticle attraction is $\lesssim \kT$. The colloids sediment rapidly as single particles to the bottom of our sample chambers. Bulk volume fractions of $10^{-4} \lesssim \phic \lesssim 2 \times 10^{-3}$ gave surface area fractions of $0.04 \lesssim \eta_{\rm c} \lesssim 0.8$.


To induce depletion attraction, we added sodium polystyrene sulphonate (NaPSS, $M_{\rm w} \approx \SI{e6}{\dalton}$; Sigma Aldrich, used as purchased). 
NaPSS behaves as a ideal neutral polymer in a medium with $I = 3.1\,\mathrm{M}$, while at $I = 0.15\,\mathrm{M}$ we have a good solvent.\cite{Wang1988} We therefore expect  swollen coils in our PMB. We estimated the overlap concentration as the inverse intrinsic viscosity,\cite{Colby2003} which came from  extrapolating Kraemer and Huggins plots of the measured viscosity as a function of polymer concentration,\cite{Colby2003, Linek2012} giving $c_p^\ast \approx 0.44\,\mathrm{wt}\,\%$. From this, we estimate a radius of gyration, $\rg$, of NaPSS in PMB using
\begin{equation}
    c_p^\ast = \frac{3M_{\rm w}}{4\pi \rg^3 N_{\rm A}}, \label{eq:overlap}
\end{equation}
where $N_{\rm A}$ is Avogadro's constant, giving $\rg \approx \SI{45}{\nano\meter}$, and a colloid:polymer size ratio of $\xi \approx 0.06$.

The contact depletion attraction, $U_0$, is proportional to the polymer activity, $a_p$, which in turn scales as the polymer concentration in a reservoir in osmotic equilibrium with the colloid-polymer mixture.\cite{cptheory, Lekker2011} In our system, this is well approximated by the bulk polymer concentration. This we report as a polymer volume fraction, $\phip$, estimated using a coil volume of ${4\pi}\rg^3/3$, so that overlap, \Eqref{eq:overlap}, corresponds to $\phip = 1$. We work in the range $0 \leq \phip \leq 0.4$.


We sealed samples into  \SI{400}{\micro\meter}-high glass capillaries and aged them for $\approx \SI{2}{\hour}$ before video recording in a Nikon Ti-Eclipse inverted microscope, typically using a $\times 50$ objective to resolve single particles. We tracked particles using a Mikrotron high-speed camera (MC 1362) to determine bulk and near-wall diffusivities. A Hamamatsu Orca 4.0 CMOS camera was used to identify clusters, from which we obtained information on nearest neighbors, defined as all of a particle's neighbors whose centres are within $2R + 2\rg + \SI{0.1}{\micro\meter} \approx \SI{1.7}{\micro\meter}$, where the \SI{0.1}{\micro\meter} accommodates polydispersity. We also used a $\times 10$ ($\mathrm{N.A.} = 3$) objective to obtain images in which single particles were not resolved. We thresholded these images, identified clusters and calculated their areas using custom software. Results for small clusters from the two methods agree up to a scaling constant between cluster area in (pixel)$^2$ and particle number. The calibrated low-resolution method gives better statistics, especially for larger clusters.

\section{Theory} \label{sec:theory}
Previously, a mean-field van der Waals free-volume theory was used to predict the wall freezing transition in a grand canonical system.\cite{Dinsmore1997} In our semi-grand canonical ensemble, \Figref{fig:ensemble}(b), the approach simplifies considerably, as there is no need to equilibrate colloids in the wall layer and the bulk.  The semi-grand canonical free energy $F$ with ideal depletants is
\begin{equation}
  F=F_0-\rhod\kT\myav{\Vfree}_0, \label{eq:theory:f}
\end{equation}
where $F_0$ is the free energy of a reference system of 2D-confined hard spheres, and $\langle \Vfree \rangle$ is the volume available to depletants of bulk number density $\rhod$ averaged over the positions of the colloids in the reference system. The depletants are mutually non-interacting (ideal), and exert an osmotic pressure $\rhod\kT$ in the bulk, which acts as a reservoir. They cannot approach $\leq \rd$ to the surface of a colloid. 

For $F_0$ we use published free energies of the hard disc fluid and hexagonally-ordered crystal\cite{Zeng1990} that account well for two-dimensional freezing.  For the former (\cf\ Eq.\ (18) in \Refcite{Zeng1990}):
\begin{equation}
  \frac{F_0^{\mathrm{(fluid)}}}{N_{\mathrm{c}}\kT}=\ln \left[ \frac{4\etac}{\pi} \right] - 1 - \ln(1-\etac)+\frac{\etac}{1-\etac},\label{eq:theory:fl}
\end{equation}
where $\etac$ is the area fraction of colloids.  After the first two ideal terms, the next two terms give the excess free energy per particle after Rosenfeld.\cite{Rosenfeld1990} For the 2D crystal, we follow Hall's procedure for 3D hard-sphere crystals\cite{Hall1972} and fitted
\begin{equation}
  \frac{F_0^{\mathrm{(crystal)}}}{N_{\mathrm{c}}\kT}=\ln\left[ \frac{4\etac}{\pi} \right] - 1 + c_0 + c_1\betac-2\ln\betac\,\label{eq:theory:fx}
\end{equation}
to the 2D crystal curve in Fig.\ 2 of \Refcite{Zeng1990} Here, $\betac=4(1-\etac/\etamax)$ where $\etamax=\pi\sqrt{3}/6\approx0.907$ is hexagonal close-packing.  Least-squares fitting gives $c_0\approx3.08$ and $c_1\approx0.30$. As expected, as  $\etac\to\etamax$, $\betac\to0$ and $F_0^{\rm crystal} \to \infty$.  With \Eqsref{eq:theory:fl} and \eqref{eq:theory:fx} we find coexistence between 2D F-X coexistence for $0.670<\etac<0.732$, agreeing with \Refcite{Zeng1990}.

For the free volume we use the standard expansion,\cite{Lekker2011}
\begin{equation} \Vfree=\mathrm{const}-N_{\mathrm{c}}V_1
  +{\textstyle{\displaystyle\sum}_{\>i>j}}
  \>V_2(r_{ij})+\dots\label{eq:theory:vfree}
\end{equation}
where $V_1$ is the excluded volume for an isolated colloidal particle allowing for the overlap with the depletion layer at the wall, $V_2(r)$ is the overlap of the excluded volumes for a pair of colloids with centre-to-centre distance $r$, and we sum over all pairs of particles $\{i,j\}$.  We do not require first two terms: they only contribute constants to the pressure and chemical potential that do not affect phase behaviour in the semi-grand ensemble.  This is in contrast to the wall freezing transition in a bulk colloidal suspension where the absolute value of the colloid chemical potential is required, including the contribution from $V_1$, to solve for the bulk-wall equilibria.\cite{Dinsmore1997}

\begin{figure}[t]
\centering
  \includegraphics[width=0.4
  \textwidth]{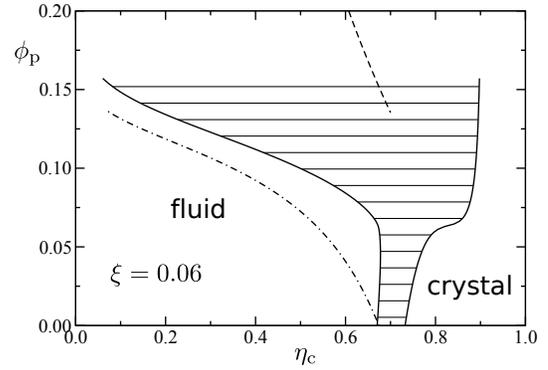}
  \caption{Phase behaviour of 2D-confined colloids in the presence of non-adsorbing ideal depletants of exclusion radius $\rd$ in the semi-grand canonical ensemble for $\xi = \rd/R = 0.06$, calculated using the approach described in the text. Horizontal tie lines span the F-X coexistence region.  The dashed line is the limit of thermodynamic stability of the fluid phase (spinodal). The chained line is the fluid binodal estimated using the simplified approach in Appendix \ref{app:coex}.\label{fig:theory}}
\end{figure}

Despite 2D colloidal confinement, the volumes in \Eqref{eq:theory:vfree} are 3D, which precludes the use of scaled particle theory to estimate $\myav{\Vfree}_0$.  However, there will be no three-body overlaps of the depletion layers of three spheres if $\rd/R = \xi < 2/\sqrt{3}-1 \approx 0.1547$,\cite{Reiss1959}. Moreover, there is no overlap of the depletion layers of two particles and the wall if $\xi < 1/4$.  Both conditions are satisfied in our case. So, we terminate  \Eqref{eq:theory:vfree} at the 2-body term and use the standard AO result
\begin{equation}
  V_2(r)=\frac{4\pi(R+\rd)^3}{3}
  \left(1-\frac{3r}{4(R+\rd)}+\frac{r^3}{16(R+\rd)^3}\right)\,.
  \label{eq:theory:v2r}
\end{equation}
To average $\Vfree$ over the reference system, we write
\begin{equation}
  \frac{\myav{\Vfree}_0}{A}=\mathrm{const}-\frac{\etac V_1}{\pi R^2}
  + \frac{1}{2}\Bigl(\frac{\etac}{\pi R^2}\Bigr)^2
  \int_{2R}^{2(R+\rd)}\!\!2\pi r\,\dr\,V_2(r)\,g(r)\label{eq:theory:int}
\end{equation}
where $g(r)$ is the radial distribution function of hard discs in the reference system, for which we we use the heuristic approximation
\begin{equation}
  g(r)=\gc+(1-\gc)\frac{r-2R}{\lambda R},\label{eq:theory:gr}
\end{equation}
where $\lambda$ captures the decay of $g(r)$ from $\gc\equiv g(2R)$ at contact.
We use $\gc=(2-\etac)/2(1-\etac)^2$ from scaled particle theory,\cite{Lebowitz1965} consistent with \Eqref{eq:theory:fl} and the sum rule for the pressure in hard disc systems.\cite{Stones2018}  For simplicity we use \Eqref{eq:theory:gr} in both the fluid and hexagonally-ordered crystal. For the fluid phase we use $\lambda=0.5$ (our results are insensitive to $0.2 \leq \lambda \leq 1$).  This corresponds to the typical contact peak in the radial distribution function in a dense fluid.\cite{Barker1976}  For the hexagonally-ordered crystal phase, $\lambda=0.1$ ensures that the effective co-ordination number is around 6 at $\etac\approx0.85$.

Substituting \Eqref{eq:theory:gr} into \Eqref{eq:theory:int} we find a dimensionless specific (\ie\ per unit area) free volume
\begin{equation}
  \frac{\myav{\Vfree}_0}{A(2R)}=\mathrm{const}-\frac{V_1\etac}{2\pi R^3}
  + \xi^3(P\gc+Q)\etac^2\,.\label{eq:theory:alf}
\end{equation}
where $Q=2\xi(1+6\xi/5+\xi^2\!/3)/3\lambda$ and $P=4(1+\xi+\xi^2\!/5)/3-Q$. Finally, combining \Eqsref{eq:theory:f} and \eqref{eq:theory:alf}, we find a dimensionless specific semi-grand free energy 
\begin{equation}  \frac{(2R)^2F}{A\kT}=\frac{4\etac F_0}{\pi N_{\mathrm{c}}\kT}
  -\rhop(2\rd)^3(P\gc+Q)\etac^2\,,\label{eq:theory:fdim}
\end{equation}
The constant and term linear in $\etac$ have been dropped from \Eqref{eq:theory:alf}, and the multiplicative factor $\xi^3$ in the final term yields the $(2\rd)^3$ which non-dimensionalises $\rhop$ in the above.

Equation \eqref{eq:theory:fdim} with \Eqsref{eq:theory:fl} and \eqref{eq:theory:fx} are solved numerically for phase coexistence as $\rhop$ increases. A typical result is shown in \Figref{fig:theory} for size ratio $\xi=0.06$.  Ideal depletants broaden the two-phase region as in the bulk,\cite{Lekker2011, cptheory} with the onset of the effect being at around $\phid=\pi\rhod(2\rd)^3\!/6 \lesssim 0.1$. We later compare this prediction to our experiments by identifying $(\rd,\phid)$ with $(\rg,\phip)$.

The limit of thermodynamic stability in the fluid can be found as the locus of points where ${\mathrm d}^2\!F/\mathrm{d}\etac^2$ vanishes.  This gives the dashed `spinodal' line in \Figref{fig:theory} (the fluid is unstable to the right of the line).  Finally, the chained line shows the prediction of the fluid binodal using a simplified approach in which we only take into account the cohesive polymer-induced AO `bond' energies in the crystal (see Appendix \ref{app:coex}).

%
%

\section{Experimental results}
\subsection{2D Confinement}\label{sec:2dconf}
At a density difference of $\Delta \rho \approx \SI{e3}{\kilo\gram\per\cubic\meter}$ with PMB, our particles sediment at $\approx \SI{1}{\micro\meter\per\second}$. A dilute suspension takes  $\gtrsim \SI{10}{\minute}$ to establish steady state in a \SI{400}{\micro\meter}-high capillary,\cite{Weaver1926} to give a height-dependent particle density (at $\phip = 0$)\cite{Poonchapter} of $n(z) = n(0)e^{-(z-R)/z_0}$, with a calculated $z_0 = \SI{0.23}{\micro\meter}$. The confinement is further increased when polymer is added to induce a depletion attraction  between the particles and the bottom capillary surface. So, we study an essentially two dimensional (2D) layer of colloids at the bottom capillary surface. 

\begin{figure}[t]
\centering
  \includegraphics[width=0.355
  \textwidth]{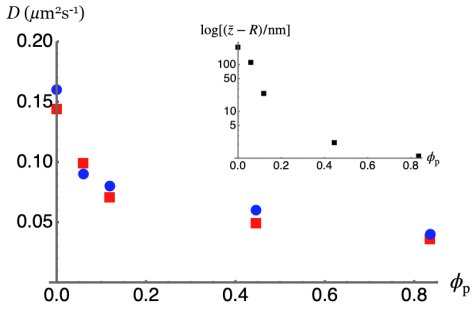}
  \caption{Diffusivity of our silica colloids next to a wall, $D$, as a function of polymer volume fraction, $\phip$. Experimental data from particle tracking ({\color{blue} $\bullet$}) and prediction using \Eqref{eq:dbar} ({\color{red} $\blacksquare$}). Inset: predicted average gap between particle (in nm) and wall as a function of $\phip$.}
  \label{fig:Dz}
\end{figure}

To quantify the confinement, we determined low $\phic$ diffusivities by tracking. The bulk diffusivity $D_0 = \SI{0.28}{\square\micro\meter\per\second}$ calculated from the Stokes-Einstein relation is reduced by near-wall hydrodynamics to 
\begin{equation}
    D = \alpha D_0,
\end{equation}
with $\alpha < 1$, \Figref{fig:Dz}. Stronger attraction reduces $D$ by lowering the particle-wall distance and so increases hydrodynamic hindrance. Fax\'en's approximate calculation predicts\cite{Faxen1921, HB73, Chio2020}
\begin{eqnarray}
    \alpha(z) = \frac{D}{D_0}  =  1 &-& \frac{9}{16}\frac{R}{z} + \frac{1}{8}\left(\frac{R}{z}\right)^3 \nonumber \\ 
    &-&\frac{45}{256}\left(\frac{R}{z}\right)^4 
    - \frac{1}{16}\left(\frac{R}{z}\right)^5. \label{eq:alpha} 
\end{eqnarray}

In the dilute limit, the probability of finding a particle at $z$ above the wall scales as $e^{-U_{\rm tot}(z)/\kT}$, where the total potential experienced by a near-wall particle is 
\begin{equation}
\frac{U_{\rm tot}^{\rm w}(z)}{\kT} = \left\{ \begin{array}{ll} 
                 \infty   & z \leq R \\
                 U_{\rm d}^{\rm w}(z)+U_{\rm g}(z)   & R < z < r + 2\rg \\
                 U_{\rm g}(z)   & R > z + 2\rg \\
                \end{array} \right. . \label{eq:Utot}
\end{equation}
Here, 
\begin{equation}
    \frac{U_{\rm g}(z)}{k_{\rm B} T} = \frac{z-R}{z_0} 
\end{equation}
is the gravitational potential, and
\begin{equation}
    \frac{U_{\rm d}^{(w)}(z)}{k_{\rm B} T} = -\left( \frac{3 \phip}{4 \rg^3} \right) R\left(2\rg - z + R \right)^2 \label{eq:udepW}
\end{equation}
is the particle-wall depletion interaction.\cite{Lekker2011} We define a dimensionless magnitude of the depletion potential at contact
\begin{equation}
    u_0^{(w)} = \left| \frac{ U_{\rm d}^{(w)}(z=R)}{k_{\rm B} T} \right| = \frac{3\phip}{\xi}. \label{eq:u0w}
\end{equation}
The near-wall average diffusivity should be given by
\begin{eqnarray}
    \bar D = \frac{D_0 \int_0^\infty \alpha(z) e^{-U_{\rm tot}^{\rm w}/\kT} \,{\rm d}z}{\int_0^\infty e^{-U_{\rm tot}^{\rm w}/\kT} \,{\rm d}z} \times \frac{\mu_0}{\mu_{\rm p}}, \label{eq:dbar}
\end{eqnarray}
taking into account the increased viscosity of the polymer solution, $\mu_{\rm p}$, relative to that of PMB, $\mu_0$, both of which we measured using standard rheometry. 

Equation \eqref{eq:dbar} gives a good account of our data, \Figref{fig:Dz}. The calculated average gap $\bar z - R$ between a particle and the surface, where $\bar z$ is obtained from \Eqref{eq:dbar} with $\beta(z)$ replaced by $z$, drops from \SI{0.23}{\micro\meter} at $\phip = 0$ to \SI{2.5}{\nano\meter} at $\phip = 0.836$, \Figref{fig:Dz} (inset). Our system, especially at $\phip>0$, is indeed  highly confined to a thin 2D layer, with the smallest gap being comparable to the expected roughness of our particles.\cite{Rentsch2006, Valmacco2016} This exercise also validates the use of the AO expression under our conditions for the polymer-induced attraction between two spherical particles, for which \Eqref{eq:udepW} is a special case (taking one sphere to have infinite radius). 


\begin{figure}[t]
\centering
  \includegraphics[width=0.45\textwidth]{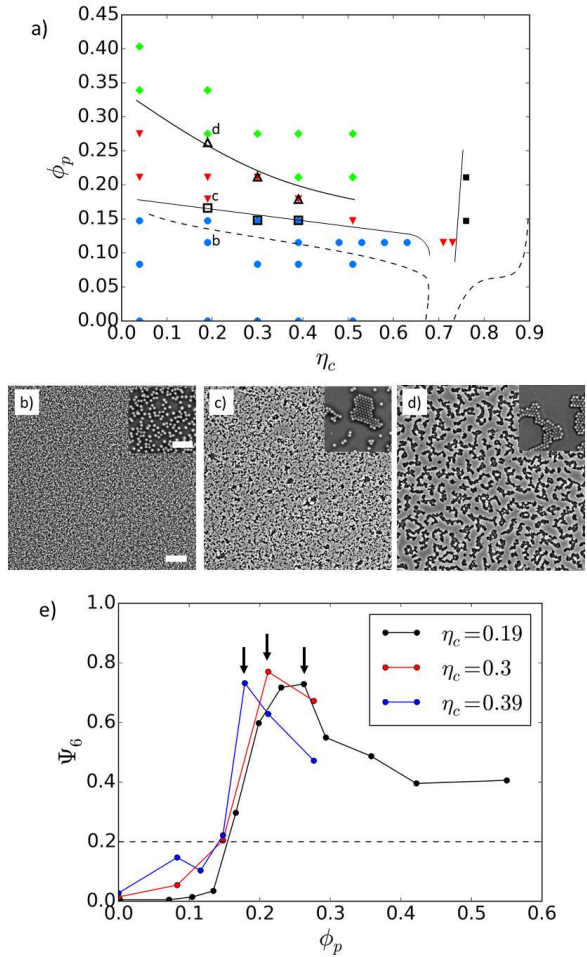}
  \caption{(a) Equilibrium phase diagram of our system. (b) Single-phase fluid, where the scale bar represents 50 $\rm{\mu m}$ and the inset scale bar represents 10 $\rm{\mu m}$. (c) Fluid-crystal coexistence. (d) Ramified cluster with crystalline local structure. (e) The bond orientation order parameter, $\Psi_6$, as a function of polymer volume fraction, $\phip$ at three different values of $\eta_{\rm c}$ (see legend). At each $\etac$, three kinds of behavior are delineated, demarcated by where $\Psi_6$ rises sharply, e.g. as quantified by where it reaches 0.2 (dashed line), and by where $\Psi_6$ peaks (arrows).}
  \label{fig:phdiag}
\end{figure}

\subsection{Phase diagram and cluster statistics}
Visually, our system shows four regimes, \Figref{fig:phdiag}(a). At $(\phip \lesssim 0.15,\eta_{\rm c} \lesssim 0.6)$, we find single particles and transient cluster with $n \lesssim 4$ particles, \Figref{fig:phdiag}(b). This is a colloidal fluid. At higher $\eta_{\rm c}$ and $\phip$. we observe F-X coexistence of single particles and colloidal crystallites, \Figref{fig:phdiag}(c). At the highest $\phip$ and $\eta_{\rm c} \lesssim 0.75$, we observe ramified clusters whose strands are crystalline, \Figref{fig:phdiag}(d). Finally, at $\eta_{\rm c} \gtrsim 0.75$, we observe a polycrystalline monolayer over a small range of $\phip$. 

We quantified the degree of crystallinity of our system via the bond orientation order parameter 
\begin{equation}
    \Psi_6 \equiv \bigl\langle|q_6(i)|^2\bigr\rangle.
\end{equation}
This is the all-particle average of the squared single-particle bond orientation parameter, which, for particle $i$ and its set of $N_i$ nearest neighbours, is given by
\begin{equation}
    q_6(i) = \frac{1}{N_i}\sum_{j =1}^{N_i} e^{i6\theta_{ij}},
\end{equation}
where $\theta_{ij}$ is the angle between the centre-to-centre vector from particle $i$ to $j$ and an arbitrary fixed axis. Perfect crystallinity gives $\Psi_6 = 1$. 

Consider first the data for $\eta_{\rm c} = 0.19$, \Figref{fig:phdiag}(e). At $\phip = 0$, $\Psi_6 = 0$. Increasing the polymer concentration, we find $\Psi_6$ remaining low until $\phip\approx 0.14$, whereupon $\Psi_6$ increases sharply, evidencing transition to F-X coexistence. The same behavior occurs at $\etac = 0.30$ and 0.39. We take the transition to F-X coexistence in each case to be the first data point where $\Psi_6 \geq 0.2$. These points, open squares in  \Figref{fig:phdiag}(a), agree well with the phase boundary determined by inspecting micrographs. 
At each $\etac$ studied, $\Psi_6$ reaches at sharp peak of $\lesssim 0.8$, and then falls. This is the onset of progressively more ramified crystalline clusters with increasing fraction of edges. We argue below that at long times, these clusters will percolation to form a gel. We therefore take the peak position in $\Psi_6$ as the gel boundary, open triangles in \Figref{fig:phdiag}(a), which again agrees with the visually-demarcated onset of ramified clustering.

\begin{figure}[t]
\centering
  \includegraphics[width=0.4\textwidth]{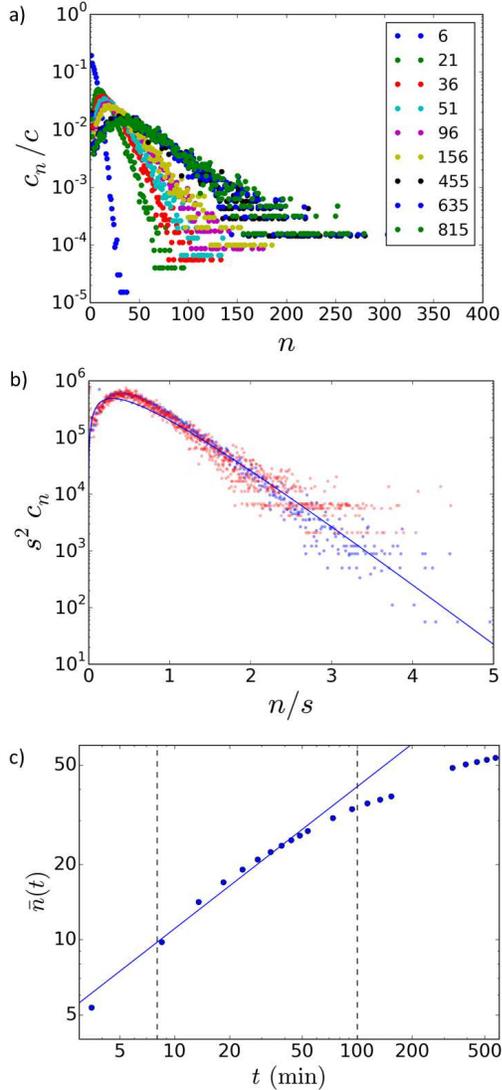}
  \caption{Cluster statistics at $\eta_{\rm c} = 0.2$ and $\phip = 0.326$, which is in the ramified clusters regime. (a) Raw normalised cluster size distribution, $c_n(t)$, at different times as in legend (in seconds). (b) Scaled normalised cluster size distribution according to \Eqref{eq:ansatz}. Blue = data sets for $\SI{10}{\minute} < t < \SI{100}{\minute}$ Red = later data sets. Continuous curve = fitting of blue data sets to \Eqref{eq:f}, giving $\lambda = -0.76$. (c) Evolution of average cluster size with time. The line has slope $z = (1-\lambda)^{-1} = 0.568$. The blue data in (b) are from the interval between the vertical lines.} \label{fig:cn}
\end{figure}

In the ramified cluster regime, we measured $c_n$, the number of clusters of size $n$.  Figure~\ref{fig:cn}(a) shows how $c_n$ normalised by the total number of cluster $c$ (so that $\sum_n c_n/c = 1$) evolves with time for a sample with $\etac = 0.2$ and $\phip = 0.326$. From these cluster size distributions (CSDs), we extracted the average cluster size $\bar n(t)$, \Figref{fig:cn}(c). It is clear that the clusters are  growing throughout our experimental time window, albeit with a reducing growth rate, suggesting that the system will percolate at even longer times to form a gel. Our experimentally-identified transition from F-X to ramified clusters is therefore the system's gel boundary. 


\section{Discussion}
\subsection{Equilibrium phase behavior}
Apart from a difference in size ratio, 
our system should be directly comparable to the simulations by Cerd\`a \etal,\cite{Cerda2004} who studied nearly-hard particles at $\eta_{\rm c} = 0.157$ interacting via an AO attractive potential with dimensional range $\xi = 0.1$ (we have $\xi = 0.06$). At $U_0 = 3.130\,\kT$, they observed a transition from `small fluctuating clusters' -- our fluid phase -- to large hexagonal close packed clusters in a background of single particles -- our F-X coexistence. Our transition from fluid to F-X coexistence occurs at $\phip \lesssim 0.16$,  corresponding to $U_0^{\rm (c)} \lesssim 4\kT$, where $U_0^{\rm (c)}  = {U_0^{\rm (w)}}/{2}$ (\cf\ \Eqref{eq:u0w}) is the dimensionless colloid-colloid depletion attraction at contact.\cite{Lekker2011} 

Next, we compare our experimental phase diagram with the theory outlined in Section~\ref{sec:theory}. To do so, we need to relate the exclusion radius of the depletant, $\rd$, to a property of the polymer in our system. Clearly, $\rd = c\rg$ for some dimensionless constant $c$. The result for taking $c=1$, so that $\phid = \phip$, is plotted in \Figref{fig:phdiag}(a). (Taking $c=1$ means we reproduce \Figref{fig:theory} in \Figref{fig:phdiag}(a).) The theory gives a credible account of the shape of the F-X boundary for $\etac < 0.67$ (where hard disks freeze). Quantitatively, the agreement is much better than order of magnitude.
Such agreement is significant. Our theory predicts the onset of significant depletion effects on the F-X transition occurs at $\rhop(2d)^3\sim O(1)$, because it is this particular dimensionless combination that enters the free energy in \Eqref{eq:theory:fdim}. If our model incorporates wrong physics, then another length, the colloid radius $R$, may also enter into the non-dimensionalisation of $\phip$, potentially altering the predicted phase boundary by a factor of $\xi$, $\xi^2$ or $\xi^3$. Given that our $\xi \sim O(10^{-1})$, the good agreement we find without fine tuning $c$ confirms that  depletion correctly captures the essential physics of the phase behavior in our system.



It is interesting to compare our phase diagram with the 3D case. We do so via the second virial coefficient, $B_2$, which is often used to compare results for potentials of different shapes in the same spatial dimension.\cite{Noro2000} In our case, we normalise $B_2$ by the hard-sphere or hard-disk values in 3D and 2D to give a dimensionless $b_2$.\footnote{We calculated $b_2$ by numerical integration using an approximation to the AO potential valid for $\xi \lesssim 0.1$ taken from \Refcite{Bergenholtz2003}} The F-X coexistence boundary for a 3D colloid-polymer mixture at $\phi_{\rm c} < 0.494$ (the onset of bulk crystallization) occurs at $b_2^{\rm (3D)} \approx -0.67$, which is also the crystallization threshold for many globular protein solutions.~\cite{Poon1997} Interestingly, our F-X boundary at $\phip \approx 0.16$, \Figref{fig:phdiag}(a), corresponds to $b_2^{\rm (2D)} \approx 0.72$. Backing out an equivalent $\phip = 0.21$ for the $\xi = 0.1$ system of Cerd\`a \etal\ via $U_0^{\rm (c)} = \frac{3\phip}{2\xi}$, we find that their F-X transition occurs at $b_2^{\rm (2D)} = 0.74$. This agrees with our experimental value, but differs significantly from the 3D value of $\approx -0.67$. Thus, in 2D, the depletion attraction does not have to be as strong to bring the system into F-X coexistence. In this sense, crystallization is easier in 2D.

Finally, we note that Li \etal\ reported an experimental study of a system of sedimented PMMA colloids ($2R_{\rm L} = \SI{3.27}{\micro\meter}$) with depletion attraction induced by smaller ($2R_{\rm S} = \SI{0.192}{\micro\meter}$) PMMA colloids (giving $\xi = 0.05$).\cite{Li2019} In contrast to our experiments and theoretical prediction, where the F-X coexistence region expands rather suddenly at $\phip \sim 10^{-1}$, \Figref{fig:phdiag}(a), their F-X coexistence  expands gradually from the $\phip = 0$ region. We do not at present understand the origins of this difference, and will not discuss their findings further.

\subsection{Gelation}
Strikingly, our ramified clusters are locally crystalline, \Figref{fig:phdiag}(d). This contrasts with 3D gelation, in which the ramified clusters are locally amorphous, displaying a variety of `locally-favored structures' such as icosahedra that are denser than crystalline packing but cannot tessellate space.\cite{Royall2008, Griffiths2017} The absence of such structures in 2D is the origin of locally-crystalline ramified clusters in our system; it is presumably also related to why crystallization is `easier' in 2D. 

In 3D, gelation in colloid-polymer mixtures is triggered by a coarsening spinodal gas-liquid phase separation texture that kinetically arrests. In our system, the theoretical fluid spinodal (dashed line in \Figref{fig:theory}) occurs far above the experimental gel boundary, \Figref{fig:phdiag}(a).  However, the spinodal is where density fluctuations diverge. At this point, the the mean field-van der Waals approximation implied by \Eqref{eq:theory:f} that underpins the free volume theory is questionable since the two terms in \Eqref{eq:theory:f} are likely to be of the same order of magnitude.  We therefore cannot rule  out that the onset of ramified structures is still  co-incident with an underlying spinodal-like thermodynamic transition in the fluid.\footnote{We note though that the free volume approach \emph{can} be justified for the F-X binodal calculation since there the effect of free volume overlaps in the fluid phase is still small, and density fluctuations are unimportant in the near-close-packed hexagonally-ordered crystal; indeed this motivates the approximate approach described in Appendix \ref{app:coex} which already yields a semi-quantitatively accurate prediction for the fluid binodal.} 

Cerd\`a \etal\cite{Cerda2004}~found ramified, locally-crystalline clusters at $U_0 = 7\,\kT$ at $\eta_{\rm c} = 0.2$, which is consistent with our data, \Figref{fig:phdiag}(a), where $\phip = 0.25$ corresponds to a contact attraction of $6.25\kB T$. Their CSD data collapsed according to the \ansatz
\begin{equation}
c_n(t) = M_1 [s(t)]^{-2} f\left[ n/s(t) \right], \label{eq:ansatz}
\end{equation}
where $M_k = \sum_n n^k c_n$ is the $k$-th moment of the CSD, and $s = M_2/M_1$ is a measure of the (time-dependent) average cluster size. We attempt this scaling for our data, \Figref{fig:cn}(b). The resulting $f$ for the earliest time, $t = \SI{6}{\minute}$ ($\approx 36 \times$ the time for a single particle to diffuse its own diameter), does not show the same peaked behavior as data from all other times. We   exclude these data from further consideration. Equation~(\ref{eq:ansatz}) collapses data from $\SI{10}{\minute} \lesssim t \lesssim \SI{100}{\minute}$, into a universal $f$ that is peaked, \Figref{fig:cn}(b) [blue]. Data from $t \gtrsim \SI{100}{\minute}$, \Figref{fig:cn}(b) [red], are increasingly noisy, and show systematic deviations from a single universal curve, especially at $n/s \gtrsim 2$. 

For DLCA, the function $f$ in \Eqref{eq:f} takes the form\cite{Dongen1985}
\begin{equation}
    f(x) = Ax^{-\lambda}e^{-ax}, \label{eq:f}
\end{equation}
where $\lambda$ is the homogeneity exponent, defined such that if $dc_n/dt = \half\sum_{i+j=n}A(i,j)c_jc_i - \sum_{j=1}^{\infty} A(n,j)c_nc_j$ (a  Smolouchowski equation), then $A(mi,mj) = m^\lambda A(i,j)$. We fitted this form to the red data in Fig.~\Figref{fig:cn}(b)to obtain $\lambda \approx -0.76 \pm 0.05$ (and $A = 2.65 \pm 0.2 \times 10^6$, $a = 2.58 \pm 0.05$).\footnote{Our fitting was done using the non-linear fitting function of Mathematica, which returns fitted parameter values and their estimated errors.}

Kinetic scaling arguments by Kolb\cite{Kolb1984} predict that the mean aggregation number in $d$-dimensional DLCA grows with time as $\bar n(t) \sim t^z$, where the dynamic critical exponent $z = \df/(\df - (d-3))$, and $\df$ is the fractal dimension of the ramified clusters.  Importantly, it can be shown that
\begin{equation}
  z = \frac{1}{1-\lambda}, \label{eq:zlambda}
\end{equation}
so that $\lambda = (d-3)/\df$. For two-dimensional DLCA,\cite{Meakin1983, Kolb1983}  $\df \approx 1.4$ implies $\lambda \approx -0.71$, which is close to our fitted value.

The above assumes that the cluster mobility  is inversely proportional to the cluster radius (\ie\ $\alpha =-\df^{-1}$ in Eq.\ (1) in Kolb) but there are reasons to believe that the far-field hydrodynamic interactions may be screened for wall-bound clusters (see Appendix~\ref{app:HI}) which would instead make the cluster mobility ultimately inversely proportional to the aggregation number ($\alpha =  -1$ in Kolb) yielding $z = {1}/{2}$ and $\lambda = -1$.  The best fit of \Eqref{eq:ansatz} with the constraint $\lambda = -1$ is almost visually indistinguishable from the unconstrained best-fit $\lambda = -0.76$ on the scale of \Figref{fig:cn}(b). Our data therefore cannot distinguish between these two models.

Our best-fit value $\lambda = -0.76$ differs from Cerd\`a \etal\cite{Cerda2004} who find $\lambda = -0.35$, but this exponent is sensitive to details such as the system concentration. More importantly, we should seek internal consistency in the form of \Eqref{eq:zlambda}. For $\lambda = - 0.76$, we expect $z = 0.568$. This dynamical exponent gives a reasonable account of our intermediate-time $\bar n(t)$ data, \Figref{fig:cn}(c). On the other hand, it is clear that our data will not be able to decide between this exponent and the value of 0.5 expected with near-wall hydrodynamic screening. 

Systematic deviations from a pure power-law behavior occurs at the end of the intermediate time window and beyond, \Figref{fig:cn}(c).
As time went on, particles increasingly adhered to  the capillary surface, especially at higher $\phip$. This is not surprising, considering the small particle-wall gaps inferred from diffusivity, \Figref{fig:Dz} (inset). The probability of adesion increases with cluster size $n$. An adhered cluster can no longer diffuse translationally, and has, at best, restricted rotational diffusivity. Such adhesion will cause deviations from either of the predicted modes of dynamical scaling. 

\section{Summary and concluding remarks}
We have studied experimentally a layer of colloids at the bottom of a glass capillary in the presence of smaller polymers. The combination of gravitational sedimentation and the depletion attraction induced between the spheres and the wall tightly confined the spheres to a 2D layer, in which we have been able to deduce the sphere-wall gap from fitting measured diffusivities to a hydrodynamic theory.\cite{Faxen1921, HB73, Chio2020} 

The depletion attraction between the spheres induced by the polymers drives them into F-X coexistence at colloid concentrations $\etac$ very much lower than that needed for 2D crystallization ($\etac = 0.67$) in the absence of polymers. A free-volume theory\cite{Poon1994} adapted to our semi-grand canonical system gives a good account of the F-X coexistence boundary.

At high polymer concentration, the depletion attraction drives the formation of ramified clusters that are locally crystalline, confirming a previous simulation.\cite{Cerda2004} However, the cluster size distribution and cluster growth dynamics show quantitative differences with these simulations, which we speculate as due to particle adhesion and/or the screening hydrodynamic interactions by the wall. Future simulations may explore the validity of these proposals.

The 500-word 1954 letter by Asakura and Oosawa published in this journal\cite{Ramos2020} together with Vrij's later detailed treatment based on the AO picture\cite{Vrij1976} marked the start of modern research into depletion-driven phenomena. A first theoretical account of the phase behavior in colloid-polymer mixtures integrates out the polymeric degrees of freedom and uses an AO potential for the inter-particle interaction.\cite{Gast1983}  Later, a `primitive model' that takes explicit account of the polymers' centre-of-mass degrees of freedom successively predicts polymer partitioning in coexisting phases.\cite{cptheory} Its applicability was confirmed by bulk experiments.\cite{cpphaseexpt} Interestingly, this primitive model uses a semi-grand canonical ensemble\cite{Lekker1990} as a calculational device. 

It is gratifying that seven decades from `AO' and three decades from the semi-grand canonical model, we are able to perform experiments in a well-characterized ensemble of this kind, fit our dynamical (diffusion) data by appealing to an AO form of the interaction between particles and wall, \Figref{fig:Dz}, and account for the equilibrium phase behavior using a modified version of the original primitive model, \Figref{fig:phdiag}(a). 

\acknowledgements
SEG was funded by an EPSRC studentship. NK, TV and WCKP were funded by ERC Advanced Grant ERC-2013-AdG 340877-PHYSAP. The data that support the findings of this study are openly available in [to be inserted].

\appendix

\section{Approximate estimate of fluid binodal}\label{app:coex}
In the presence of non-adsorbing polymers the cohesive free energy of the hexagonally-ordered crystal can be approximated by calculating the energy required to break the AO `bonds', as in a solid-state physics problem.\cite{Kittel}  By matching this to the colloid chemical potential in the fluid phase, one can estimate of the location of the fluid binodal.  Let the crystal co-ordination number be $z$.  Then
\begin{equation}
  F^{\mathrm{(crystal)}}\approx F_0^{\mathrm{(crystal)}}-
  \frac{zN_{\mathrm{c}}}{2}\times\rhop\kT\, V_2(2R)
\end{equation}
where the second factor in the second term is the AO bond free energy, and for simplicity we use the contact value of $V_2(r)$.  Equating the resulting colloid chemical potential to that in the two-dimensional fluid derived from \Eqref{eq:theory:fl} gives
\begin{equation}
  \ln\left( \frac{\etac}{1-\etac} \right)+\frac{\etac(3-2\etac)}{(1-\etac)^2}=\frac{\mu_0}{\kT}-
  \frac{z}{2}\times \rhop V_2(2R)\label{eq:app:theory:mu}
\end{equation}
where $\mu_0$ is the chemical potential in the unperturbed crystal, which in the spirit of the approach we shall suppose constant.

For simplicity we neglect the excluded volume overlaps in the fluid (one can show that they are small) and tacitly omit the $V_1$ terms which contribute only a common constant to the chemical potentials.  Since at coexistence the chemical potentials of the fluid and crystal are the same, $\mu_0$ can be obtained from the known fluid coexistence composition in the absence of added polymer, \viz\ \Eqref{eq:app:theory:mu} should be verified by $\etac\approx 0.670$ at $\rhop=0$.  Solving \Eqref{eq:app:theory:mu}  (with $z=6$) for $\etac$ as a function of $\rhop$ then provides an estimate of the fluid binodal, shown for the present system as the chained line in \Figref{fig:theory}.

\section{Hydrodynamic interactions} \label{app:HI}
We present heuristic arguments that in a wall-bound cluster containing $N$ particles, the wall `screens' the hydrodynamic interactions such that the cluster mobility $\sim1/N$, at least in the scaling limit.  We start with the familiar result that a point force $\fvec$ in an unbounded fluid generates a velocity field $\vvec$ at a distance $\rvec$ with (Oseen tensor)\cite{Doi86}
\begin{equation}
  \vvec=\frac{1}{8\pi\mu r}
  (\Ivec+\rhatvec\,\rhatvec)\cdot\fvec\,.\label{eq:app:hi:oseen}
\end{equation}
In this $\Ivec$ is the unit tensor, $\rhatvec=\rvec/r$ with $r=|\rvec|$, and $\mu$ is the fluid viscosity.  Similarly, Blake and Chwang showed that a point force at a height $h$ above the wall generates a flow field which behaves to leading-order in far-field as\cite{Blake74}
\begin{equation}
  \vvec=\frac{12hz}{8\pi\mu r^3}\>
  \rhatvec\,\rhatvec\cdot\fvec\label{eq:app:hi:vwall}
\end{equation}
Here, the no-slip boundary condition is coincident with the $z=0$ plane, and $\rvec=(x,y)$ is now the in-plane distance between the point $(x, y, z)$ where the velocity is measured and the point $(0, 0, h)$ where the force is applied.  Crucially, according to \Eqref{eq:app:hi:vwall}, for $z\sim h$ the far-field decays as $1/r^3$ rather than $1/r$ as in \Eqref{eq:app:hi:oseen} for an unbounded fluid.\cite{note-asymp}  

At this point we recall that the mobility of a fractal cluster is essentially determined by the behavior of $\myav{1/r_{ij}^{\>n}}$ where the average is taken over all pairs of particles in the cluster,\cite{Wiltzius1987} and according to the above we should take $n=1$ for freely-suspended clusters or $n=3$ for wall-bound clusters.  In terms of the pair distribution function $g(r)\sim r^{\,\df-d}$,
\begin{equation}
  \myav{1/r_{ij}^{\>n}}
  \sim\frac{{\textstyle\int}\, r^{-n}\,g(r)\,r^{\,d-1}\,\dr}
  {{\textstyle\int}\,g(r)\,r^{\,d-1}\,\dr}
  \sim R^{-n}\label{eq:app:hi:rav}
\end{equation}
where $R\sim N^{1/\df}$ is the cut-off in $g(r)$.  But this \emph{only} holds when the integral in the numerator is dominated by this cut-off, which requires $\df>n$.  Plainly this is the case for freely-suspended clusters (unless they happen to be fractal dust with $\df<1$), and so one expects that the cluster mobility $\sim N^{-1/\df}$.  This scaling behavior has been widely confirmed,\cite{VanSaarloos1987, Warren1994, Sorensen2011} and corresponds to the fact that the flow field is screened from the interior of the cluster.  For wall-bound clusters though, it is \emph{not} the case that the integral in the numerator in \Eqref{eq:app:hi:rav} is dominated by the upper limit since that would require $\df>3$ which is impossible in $d=2$ dimensions.  Hence one concludes (perhaps a little tentatively!) that hydrodynamic interactions should be negligible in far-field for wall-bound clusters, or in other words the flow field, already screened by the wall, is not further significantly reduced in the interior of the cluster.  It follows that the frictional drag should be extensive in the number of particles, and the cluster mobility should scale as $\sim1/N$ as claimed.

This conclusion obviously demands numerical verification, using Stokesian dynamics or similar methods.\cite{Cichocki2000, Swan2007, Gauger2009}  For now though, we close with a couple of remarks higlighting the subtleties of this hydrodynamic problem.  First, it is interesting to note that the mobilities of individual particles are in a sense \emph{decoupled} from the hydrodynamic interactions, since the former are sensitive to the gap between the particles and the wall (see section \ref{sec:2dconf}), which can be made arbitrarily small, whereas the latter are essentially controlled by the heights of the particle centres above the wall which are limited by the particle radii.  Second, whilst the presence of the wall couples the rotational and translational modes, in a cluster the rotational modes are partly supressed if the particles are mutually hindered from all rolling in the same direction (essentially as a non-trivial consequence of the near-field hydrodynamics).  So, it seems doubtful that the individual particle friction coefficients are simply additive, but this may not necessarily change the extensivity of the overall cluster drag coefficient. 

\bibliography{depletion}

\begin{thebibliography}{54}
\bibitem [{not()}]{note-asymp}%
  \BibitemOpen
  \href@noop {} {}\bibinfo {note} {\Eqref{eq:app:hi:vwall} is derived from the
  Blake tensor in the limit $r\to\infty$ at finite $z$, and differs from the
  better-known asymptotics for the flow field \emph{distant} from the wall
  (\ie\ $r, z\to\infty$ with $z/r$ held constant) for which
  $|\vvec|\sim1/r^2$.}\BibitemShut {Stop}%
\end{thebibliography}

\end{document}